\documentclass{aa}
\usepackage{graphicx}
\usepackage{txfonts}
\usepackage{amssymb}
\usepackage{lscape}
\usepackage[dvips]{epsfig}
\begin{document}

\title{Limb polarization of Uranus and Neptune \thanks{Based on observations obtained at the ESO 3.6m Telescope at La Silla, Chile (ESO program 72.C-0498).}}

\subtitle{\sc {II}. Spectropolarimetric observations}

\author{F.~Joos
  \and H.M.~Schmid}

\offprints{Franco Joos, \email{fjoos@astro.phys.ethz.ch}}
              
\institute{Institut f\"ur Astronomie, ETH Z\"urich, 8092-Z\"urich, Switzerland}

\date{Received / Accepted  }

\abstract
{}
{We have detected a strong limb polarization for Uranus and
Neptune. With spectropolarimetric observations we characterize
the spectral dependence of this limb polarization and explore the diagnostic potential for investigating
the distribution and properties of the
scattering particles.} 
{We present disk resolved spectropolarimetry of Uranus and Neptune
covering the wavelength range from 5300\,\AA\ to 9300\,\AA\ and
compare the spectropolarimetric signal for different limb sections
and the center of the planetary disk. As an additional benefit we obtained
center-to-limb disk profiles for the intensity
and polarization for various wavelengths.}
{Our data show for both planets a strong linear limb polarization
oriented perpendicular to the limb. The polarization at the center
of the planetary disks is essentially zero.
Overall the limb polarization increases towards shorter wavelengths, and 
it is further enhanced in the methane
absorption bands when compared to adjacent inter band regions. 
Qualitatively, the polarization signal
is very similar for Uranus and Neptune. 

For Uranus the peak limb polarization of the methane absorption
bands reaches a  maximum of 3\,\% whereas in the nearby higher albedo regions
it peaks at about 2\,\%. The measured polarization for Neptune is
significantly lower because the signal is averaged down by the
limited spatial resolution of our Neptune data. 

The disk profiles for Uranus (center to limb
profiles) show for the intensity a strong asymmetry between the bright South pole regions and the darker northern latitudes. 
The asymmetry is particularly strong in the methane bands. 
We measure a significant limb brightening in the methane bands at the Uranian eastern and western limbs. The intensity asymmetry between North and South and the limb brightening display a tight 
anti-correlation with the albedo of the selected wavelength interval.

From the polarization profiles of Uranus we determine slit integrated
polarization values in order to derive intrinsic limb polarization parameters 
which are corrected for the limited spatial resolution of our data. 
These parameters are compared with previous imaging polarimetry.}
{The polarimetric measurements of Uranus and Neptune are compatible 
with expectations for the limb polarization of
scattering atmospheres containing substantial layers of particles which 
scatter radiation similar to Rayleigh scattering. It seems
that the limb polarization contains interesting diagnostic information,
in particular on the scattering properties of the
aerosol particles in the haze layers at high altitudes.}


\titlerunning{Limb polarization of Uranus and Neptune. \sc{II}}

\maketitle

\section{Introduction}
In a previous paper we have reported the detection of limb polarization of about 1\,\%
for Uranus and Neptune with imaging polarimetry in the R, i and z bands 
(Schmid et al.~\cite{schmid06}; hereafter referred as paper {\sc i}). 
The linear limb polarization was found to be perpendicular to the limb 
and roughly equally strong along the entire limb.
This is unlike Jupiter where the strength of the limb polarization 
is characterized by a strong difference between polar and equatorial 
limb regions (Hall \& Riley~\cite{hall76}; 
Smith \& Tomasko~\cite{smith84}; Braak et al.~\cite{braak02}; 
Joos et al.~\cite{joos05}). 

The limb polarization can be explained as a second order scattering effect which 
is expected to occur for reflecting atmospheres seen near phase angle $\alpha=0^\circ$ 
(backscattering) where Rayleigh-type scattering processes are dominant 
(see van de Hulst~\cite{vandehulst80}). 

In this second paper we present disk resolved spectropolarimetry of Uranus and Neptune. 
Our data cover a wavelength range from 5300\,\AA\ to 9300\,\AA\ which includes 
strong methane absorption bands. Strong absorption bands correspond to strong 
opacity jumps in the scattering atmospheres and corresponding polarization
effects may be expected for the reflected light. 

In Sect.~\ref{6212obsdatred} we discuss the observations and the data reduction. 
The overall polarization properties of Uranus are described in Sect.~\ref{6212specpoluranus}, 
and in the following section we focus on the disk profiles (center to limb profiles)
for the intensity and the polarization in some
specific wavelength intervals. The spectropolarimetric properties of Neptune are
described in Section \ref{6212specpolneptune}. In Sect.~\ref{6212discussion} the spectropolarimetric data from this paper are 
compared to the imaging polarimetry presented in paper~{\sc i} and the results are summarized and discussed in the broader context of 
scattering polarization from planetary atmospheres.

\section{Observations and data reduction}\label{6212obsdatred}
Spectropolarimetric observations of Uranus and Neptune were taken on November 29, 2003 with EFOSC2 at the ESO 3.6m telescope at La Silla. 
During the same run also imaging polarimetry of Uranus and Neptune was obtained with the same instrument, but with a different setup (see paper~{\sc i}).

EFOSC2 is a multi-mode imager and grism spectrograph which can be equipped with a 
Wollaston prism and a rotatable superachromatic half-wave plate for linear polarimetry and 
spectropolarimetry. The Wollaston prism produces together with a grism two spectra 
with opposite linear polarization $I_\parallel$ and $I_\perp$ (the ordinary and extraordinary beam) offset by 20$''$ on the CCD. In order to avoid confusion a special 
slit mask is introduced at the telescope focus with a series of aligned slits 
having the lengths of 19.7$''$ and periods of 42.2$''$. 
For both planets we have taken data with the slit centered on the planet along 
the N-S direction in the celestial coordinate system. For Uranus we have also 
obtained data with slits in the E-W direction which was achieved by rotating the 
whole instrument. The spectrographic slits are much longer than the diameters of the planets which are about 3.5$''$ for Uranus and 2.2$''$ for Neptune. Thus, we obtained "long-slit" spectropolarimetry for both planets. 
The chosen slit width was 0.5$''$ for Uranus and 1.5$''$ for Neptune, respectively. 
For Neptune, a large fraction of the planet was visible through the slit 
except for the E and W limbs.

The employed grism (EFOSC2 grism\#5) covers the wavelength range of 5200 - 9350\,\AA\ and provides a spectral resolution of 12.8\,\AA\ for a 1$''$ wide slit. The data were recorded with a $2{\rm k} \times 2{\rm k}$ CCD (ESO CCD\#40) with a spatial scale of $0.157''\,{\rm pixel}^{-1}$ and a wavelength scale of 2.06\,\AA\ per pixel. 

The effective seeing for our spectropolarimetric data was about 1$''$ as derived from the width of the spectra of the standard stars. 

Our polarimetric data are expressed in form of the normalized Stokes parameters $Q/I$ and $U/I$.  There is $Q/I = (I_\parallel - I_\perp$)/($I_\parallel + I_\perp) = (I_{0} - I_{90}$)/($I_{0} + I_{90}$), where $I_\parallel = I_{0}$ and $I_\perp = I_{90}$ are the intensities with polarization direction parallel and perpendicular to the slit, respectively. The Stokes parameter $U/I = (I_{45}-I_{135}$)/($I_{45}+I_{135}$) describes polarized light in the 45$^{\circ}$/135$^{\circ}$ direction, where positive angles are rotated counterclockwise. The polarization position angles are defined relative to the slit orientation.

The linear polarization was measured in a standard way (e.g. Tinbergen \& Rutten~\cite{tinbergen92}) with sets of four exposures taken with half-wave plate position angles of $0^\circ$, $22.5^\circ$, $45^\circ$ and $67.5^\circ$, respectively. Three sets were taken for each slit orientation and both planets in order to enhance the signal-to-noise. One single exposure was 30\,sec for Uranus and 50~sec for Neptune. Polarimetric standard stars were observed with the same instrumental setup. Exposures of a helium-argon lamp provided the wavelength calibration.

Exposures with the half-wave plate angles at $0^\circ$ and $45^\circ$ yield Stokes $Q$ and the other two yield Stokes $U$. One $Q/I$ spectrum results from the combination of 4 planet spectra according to
\begin{equation}
\frac{Q}{I} = \frac{R-1}{R+1} \hspace{0.5cm} {\rm with} \hspace{0.5cm} R^{2} = \frac{I_\parallel (0^\circ)/I_\perp (0^\circ)}{I_\parallel (45^\circ)/I_\perp (45^\circ)}
\end{equation}
and equivalent for $U/I$ with the $I_{\parallel}$ and $I_{\perp}$ taken at half-wave plate angles of 22.5$^{\circ}$ and 67.5$^{\circ}$, respectively.

The polarization reduction was performed with the {\sc midas} software package similar to the description in paper~{\sc i}. Important in the data reduction procedure is that the different planetary spectra, e.g. $I_\parallel (0^\circ)$, $I_\perp (0^\circ)$, $I_\parallel (45^\circ)$ and $I_\perp (45^\circ)$, are centered in the same way and with an accuracy of about 1/10 of a pixel in spatial direction. This high precision is required because a small misalignment would produce a 
spurious polarization signal (say positive) at one limb and an opposite 
polarization signal (negative) at the other limb of the long-slit spectrum. 
For an accurate alignment it has to be taken into account that the Wollaston has 
some dispersive power. The resulting ordinary and extra-ordinary spectra are 
therefore slightly bended with a curvature which differs between the two beams. 
The alignment of the long-slit spectra of the planets was achieved with the help of the standard star observations from which the wavelength dependent spatial position of the spectra could be determined with sufficient precision. 

The polarization zero point of the instrument was derived and corrected from an unpolarized (HD 14069) and a polarized (BD $+25^\circ 727$) standard star. 
The instrumental polarization was found to be less than 0.2\,\%. 
According to the EFOSC2 documentation the instrument polarization should be less than 0.1\,\% for the central parts of the CCD (http://www.ls.eso.org/lasilla/sciops/3p6/efosc/).
The polarization angle calibration should be accurate to about $\theta\approx \pm 2^\circ$. 

The CCD introduces for $\lambda>7000$\,\AA\ an interference pattern with a full
amplitude of about 5\,\% at 7500\,\AA\ and about 10\,\% at 8200\,\AA\ of the mean intensity level. 
The differential data reduction for the polarization reduces this fringes to a full amplitude level for $Q/I$ of $\Delta p \approx 0.3\,\%$ at 7500\,\AA\ and $\approx 0.5\,\%$ at 8200\,\AA.
In the intensity and polarization spectra the fringes are visible as quasi-periodic pattern with a periodicity of about $15 - 20$\,\AA. We have tried to reduce the fringe pattern with the available flat field calibrations. Some improvement was achieved for the intensity spectrum but not for the polarization spectrum. Spectral binning turned out to be much more efficient. This averages out the periodic pattern and produced much smoother intensity and polarization spectra. As the spectral features in Uranus and Neptune are very broad the loss in spectral resolution was acceptable. With this method the spurious fringe pattern vanished almost completely and in addition a significantly 
improved S/N per bin was obtained for the polarization spectrum. 
Even with no fringes a similar spectral binning would have been necessary for an analysis of the polarization spectrum in order to have sufficient signal-to-noise. 
For these reasons we have binned all
spectropolarimetric data into 30\,\AA\ bins. 

\noindent
\begin{figure}
\begin{minipage}[t]{.45\linewidth}
\centering 
\epsfig{file=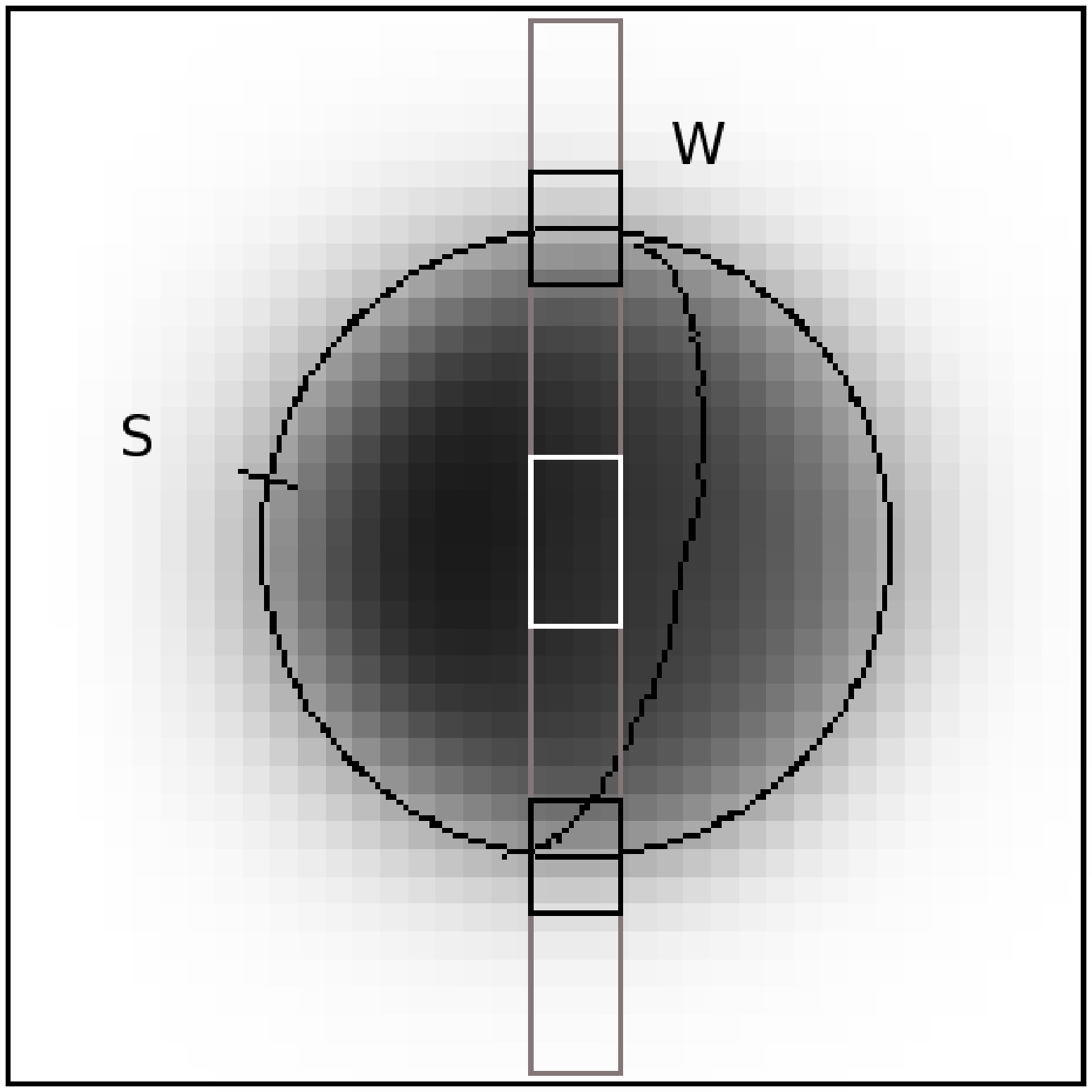,width=\linewidth} 
\end{minipage}\hfill 
\begin{minipage}[t]{.45\linewidth}
\centering
\epsfig{file=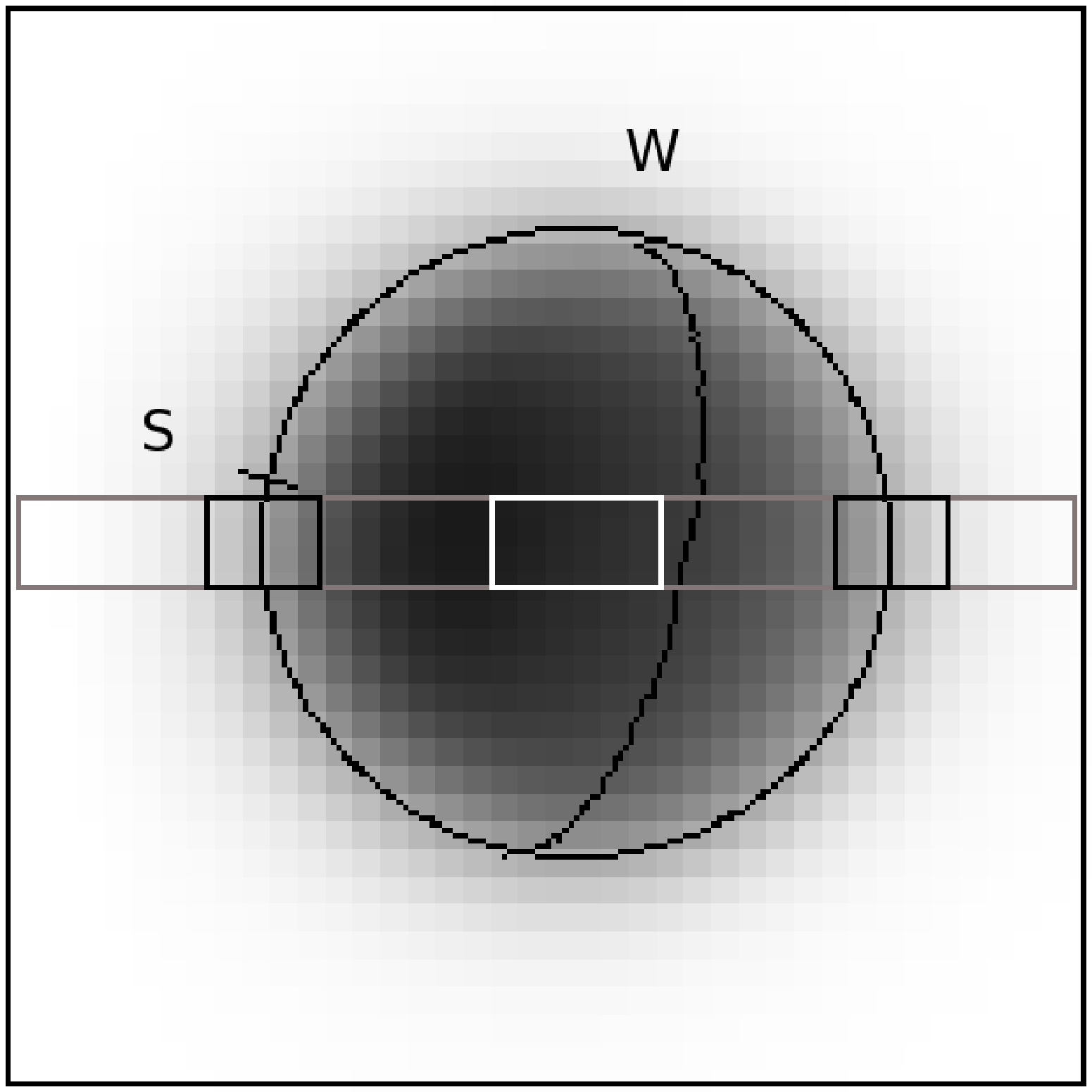,width=\linewidth} 
\end{minipage}\hfill
\caption{Uranus i-band images with the slit in celestial N-S (left) and E-W (right) orientation. N is up and E is left. The black boxes indicate the ``limb'' sections of the long-slit spectropolarimetry and the white boxes the ``central'' regions. Also shown are the positions of the South pole and the equator with the West limb indicated according to the planetary coordinates defined by Seidelmann et al. (\cite{seidelmann02}). At the time of our observations the diameter was 3.51$''$, the phase angle 2.8$^{\circ}$, the position angle of the South pole was 78$^{\circ}$ and the distance of the South pole from the disc center was 1.62$''$. (The planet parameters are taken from ``The Astronomical Almanac 2003''.)}
\label{6212fig1}
\end{figure}

\section{Spectropolarimetric structure of Uranus}\label{6212specpoluranus}
For Uranus we have taken 5300 to 9300\,\AA\ long-slit spectropolarimetry  
with the slit oriented in N-S and E-W 
direction in the celestial coordinate system (Fig.~\ref{6212fig1}). 
The N-S slit position extends roughly along the equator of Uranus. Henceforth, measurements taken with this slit position are called ``equatorial'' with the definition of West and East limbs or longitudes on the planet as indicated in Fig. \ref{6212fig1}.

The E-W slit extends from the South pole over the equator to northern latitudes roughly along the central meridian of Uranus. Measurements from this slit orientation will be referred to as ``meridional''. 
 
Figure~\ref{6212fig1} illustrates the exact slit positions and the selected regions: 
four limb regions East, West, South and North\footnote{From here on East, West, South and North refer always to the planetary coordinates.} (0.628$''$ long black boxes), two central slit regions
(0.942$''$ long white boxes), as well as the two
spectropolarimetric  signals obtained by integrating 
over 7.85$''$ along the equatorial and the meridional slit.  
These selected regions are denoted as \textit{limb}, \textit{central
  part} and \textit{total slit}. 

Already our ground based images reveal a strong asymmetry between the brighter South pole and darker northern latitudes which are well resolved as band structures in HST images (e.g. Karkoschka~\cite{karkoschka01}; Rages et al.~\cite{rages04}), but no strong longitudinal features are seen for Uranus (see also press release STScI-PRC2004-05).

\subsection{Spectroscopic structure and spectral albedo $A(\lambda)$}

\begin{figure}[h]
\centering
\epsfig{file=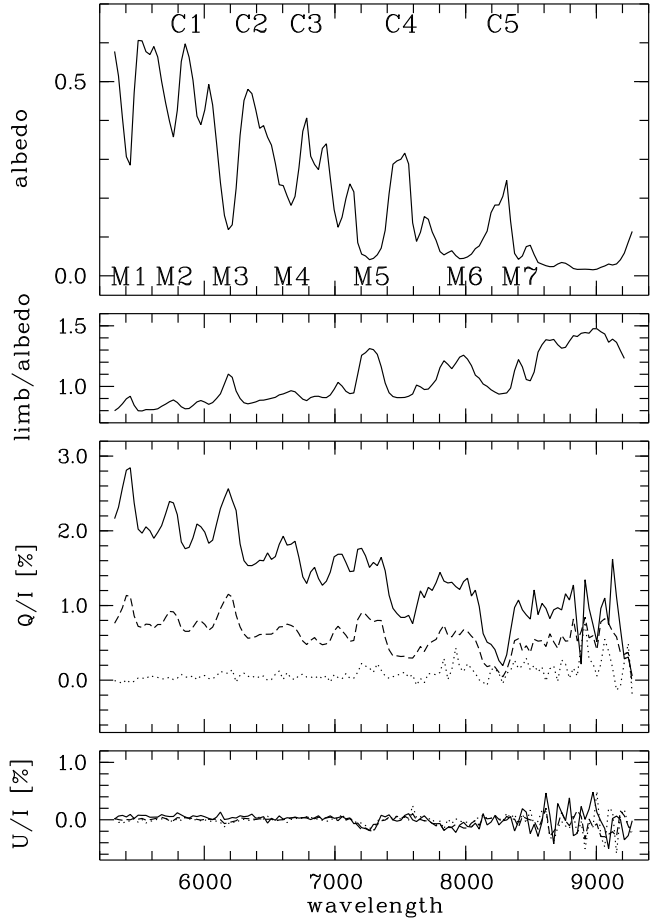,width=9cm}
\caption{Spectropolarimetry of Uranus for the equatorial slit. Top
  panel: geometric albedo spectrum $A(\lambda)$ with identifications of selected spectral
bands. Second panel: normalized ratio between the limb intensity spectrum
 and the geometric albedo spectrum $A(\lambda)$. Third panel: 
 normalized Stokes $Q/I$ polarization spectra for the average of the western and eastern limb regions (solid line), the central part (dotted line), and
 the total slit (dashed line). Bottom panel: same as third panel but for Stokes $U/I$.}
\label{6212fig3}
\end{figure}

Spectroscopically (Fig.~\ref{6212fig3}, top panel), our data show mainly the strong 
methane (CH$_4$) absorption
bands, which are well known for Uranus (e.g. Baines et al.~\cite{baines86}, 
Karkoschka~\cite{karkoschka94}). Also clearly visible is 
the telluric absorption due to the A-band at 7590\,\AA. Further, the enhanced limb brightening in the methane bands is well visible in our data and illustrated (in Fig. \ref{6212fig3}) by the normalized limb to albedo ratio.

An albedo spectrum is constructed from our data. For this we
averaged first the total slit spectra in equatorial and meridional direction and then
normalized the result to the geometric albedo spectrum of Karkoschka~(\cite{karkoschka94}) multiplying our spectrum with a smooth correction 
function which accounts for the instrumental efficiency curve. This procedure
yields a well defined full disk albedo spectrum $A(\lambda)$ as plotted
in Fig.~\ref{6212fig3}.

Spectral regions were selected in order to quantify the albedo and
polarization. The selected regions, which represent strong absorption bands 
(M1 -- M7) and the higher albedo regions (C1 -- C5) in between,
are indicated in Fig.~\ref{6212fig3} and listed in Table~\ref{6212regionsdef}. 

We found a clear difference of the spectral structure of the limb when compared
to the albedo spectrum $A(\lambda)$. This is shown in Fig.~\ref{6212fig3}
where the normalized ratio between the average of the West and East limb spectra 
and the albedo spectrum is plotted. 
The ratio is enhanced in all CH$_4$ bands indicating that 
absorption is less deep for the limb regions. This is equivalent to the
statement that deep absorption bands of Uranus show a limb brightening effect
as observed e.g. by HST (see Karkoschka~\cite{karkoschka01}). 

\subsection{Spectropolarimetry for the equatorial slit}

Spectropolarimetry for the equatorial slit may be considered as representative for the ``average'' planetary disk of Uranus
because the bright South pole region and the faint northern latitudes are not
included in these observations. The peculiarities of these special regions
are discussed in the following subsection.    

The polarimetric signal of the equatorial cut through Uranus is plotted in 
Figure~\ref{6212fig3} as normalized Stokes $Q/I$ and $U/I$ spectra 
($p_Q(\lambda)$ and $p_U(\lambda)$
respectively). 
There exists in our data no significant difference between the polarization
signal of the western and eastern limbs and therefore only the average of the two limb polarization
spectra is plotted. Fig.~\ref{6212fig3} includes further the polarization 
spectra of the central part and the average for the entire slit. 

The data show, that the $Q/I$-polarization is
positive and high at the limbs and close to zero for the central
region. Essentially no signal is visible for the $U/I$-data,
indicating that the orientation of the polarization is everywhere parallel to
the slit equivalent to a limb polarization perpendicular to the
limb. The $U/I$-spectra give a good representation of the noise errors in the data, 
which are particularly large in the deep absorption bands at the longest 
wavelength $\lambda>8500$\,\AA, where the photon statistics are poor. 

The $Q/I$-polarization spectrum for the equatorial limbs displays a clear overall 
decrease of the polarization level with wavelength. Further, the limb polarization 
is enhanced in the deep CH$_4$ absorption bands when compared to the adjacent 
continuum or spectral regions with higher albedo.

Table~\ref{6212regionsdef} gives the measured polarization for all four limbs for
selected wavelength regions. The listed values are flux weighted average values
$\langle Q/I \rangle = \Sigma Q/\Sigma I$, where $\Sigma I$ and $\Sigma Q$ are summed over intensity flux and polarization flux respectively for the selected
spatial bin and wavelength interval.

In the methane bands the limb polarization decreases from about
2.6\,\% in the band at 5400\,\AA\ to about 0.9\,\% in the broad, deep band
centered at 8900\,\AA. The measured limb polarization for the continuum (or high
albedo regions) is about 2\,\% at 5500\,\AA, and less than 0.4\,\% for the C5-region
at 8270\,\AA, except for the northern limb where the polarization for C5 is 
significantly higher.

It should be noted, that the level of polarization
in Fig.~\ref{6212fig3} and Table~\ref{6212regionsdef} depends on the seeing 
conditions and the size 
and position of the spatial bin, selected for the averaging. However, the
relative wavelength dependence of polarization for a given limb section
depends very little on 
spatial position, and therefore on the selected slit region and the seeing conditions. 
Uncertainties in the polarization values of Table~\ref{6212regionsdef} due to photon noise 
are about $\Delta \langle Q/I \rangle = \pm 0.10$\,\%, except for the narrow interval M3 and the longest CH$_{4}$ wavelength intervals M6 and M7, where 
$\Delta \langle Q/I \rangle = \pm 0.20$\,\%.

\subsection{Peculiarities for the meridional slit}
\label{6212sectEW}

Uranus, close to its equinox, which is on the 7$^{\rm th}$ of December 2007, shows currently an intensity asymmetry from the bright South pole
to the darker northern latitudes which is particularly strong
in the methane bands. This intensity asymmetry is also obvious in the HST-images of 
Karkoschka~(\cite{karkoschka01}) or Rages et al.~(\cite{rages04}) and in Keck observations obtained with an adaptive optic system (Sromovsky \& Fry \cite{sromovsky05b}).

\begin{figure}[h]
\centering
\epsfig{file=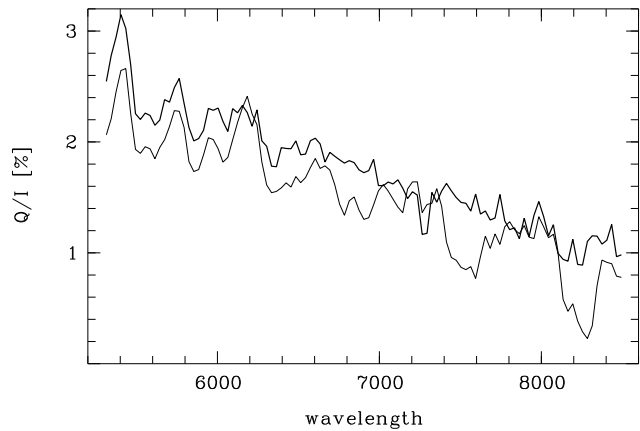,height=9cm,angle=270}
\caption{Comparison between $Q/I$-polarization spectra of the northern limb of 
Uranus (thick line) and the average of the other three limbs (thin line).}
\label{6212fig4}
\end{figure}

Also in the spectropolarimetric signal we found some clear peculiarity
for the northern latitudes. Figure~\ref{6212fig4}
compares the $Q/I$ limb spectropolarimetry of the northern latitudes
with the average of the three other limb regions, East, South and West which are very
similar and can be taken together. The northern latitudes show 
for wavelength longer than about 6000\,\AA\ no polarization minimum for
spectral regions with high albedo. Thus the normalized
polarization spectrum $Q/I$ for the northern latitude limb is essentially 
featureless and displays just a steady decrease in polarization with wavelength
(see also Table~\ref{6212fig4}).

\begin{table}[h!]
\center
\caption{Limb polarization $\langle Q/I \rangle$ for Uranus for the West, East, South and North limb and selected wavelength intervals (column 2). Column 3 gives the averaged
albedo $A(\lambda)$ for the corresponding interval.}
\label{6212regionsdef}
\begin{tabular}{ccccccc}
\noalign{\smallskip\hrule\smallskip}
feature & wavelength & albedo & \multicolumn{4}{c}{$\langle Q/I \rangle$ [\%]}  \cr
        &    [\AA]   &        & \scriptsize{West}  & \scriptsize{East} & \scriptsize{South} & \scriptsize{ North}  \cr
\noalign{\smallskip\hrule\smallskip}
C1 & 5827 - 5883 & 0.59 & 1.88 & 1.73 & 1.55 & 2.11\cr
C2 & 6300 - 6400 & 0.48 & 1.59 & 1.54 & 1.45 & 1.91\cr
C3 & 6755 - 6794 & 0.41 & 1.40 & 1.30 & 1.30 & 1.85\cr
C4 & 7442 - 7565 & 0.32 & 0.89 & 0.88 & 0.75 & 1.47\cr
C5 & 8215 - 8320 & 0.23 & 0.27 & 0.33 & 0.20 & 1.10\cr
\noalign{\smallskip\hrule\smallskip}
M1 & 5368 - 5459 & 0.37 & 2.68 & 2.63 & 2.20 & 3.10\cr
M2 & 5732 - 5788 & 0.40 & 2.41 & 2.26 & 1.96 & 2.60\cr
M3 & 6180 - 6200 & 0.13 & 2.45 & 2.52 & 1.96 & 2.29\cr
M4 & 6627 - 6699 & 0.21 & 1.84 & 1.82 & 1.51 & 1.93\cr
M5 & 7174 - 7358 & 0.06 & 1.55 & 1.70 & 1.20 & 1.40\cr
M6 & 7933 - 8034 & 0.05 & 1.27 & 1.25 & 0.96 & 1.37\cr
M7 & 8369 - 8453 & 0.06 & 0.88 & 1.03 & 0.66 & 1.21\cr
\noalign{\smallskip\hrule\smallskip}
\end{tabular}
\end{table}

According to our data, the darker northern latitudes show, when compared to the other limbs, for $\lambda <$ 6000\,\AA\ a higher overall polarization and for $\lambda >$ 6000\,\AA\ a higher polarization for the inter band regions. We have carefully investigated whether this result could be due to a misalignment or another spurious
effect. We found that the maximum error in misalignment for the polarimetric data reduction
produces an effect on the order of only 0.05. Based on such tests, we conclude 
that the described Stokes $Q/I$ polarization peculiarities of the northern limb are real.

\subsection{Fitting the limb polarization spectrum}
\label{6212sectfit}

We found that the $Q/I=p_Q(\lambda)$ polarization spectra for Uranus show a clear overall 
decrease of the polarization level with wavelength. Further, the 
polarization is enhanced in the deep CH$_4$ absorptions bands, 
when compared to the adjacent higher albedo spectral regions, 
except for the red part of the northern limb spectropolarimetry.

In order to quantify the wavelength and albedo dependence of the 
limb polarization, we fit the polarization
spectra with a simple linear relation with respect to the wavelength $\lambda$ and
the albedo $A(\lambda)$:

\begin{equation}
\label{6212fitequation}
p_Q(\lambda) [\%] = c_1\, +  c_2\cdot\lambda[\mu{\rm m}] + c_3 \cdot A(\lambda)\,. 
\end{equation} 

\begin{figure}[h]
\centering
\epsfig{file=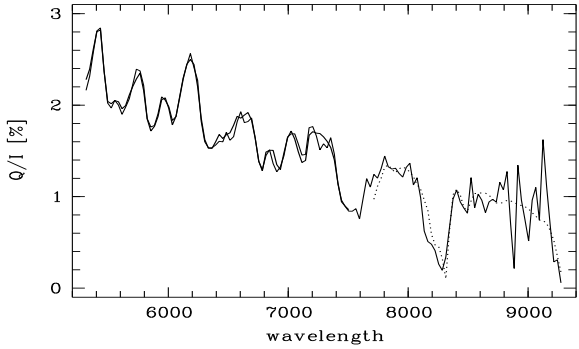,width=5cm,height=9cm,angle=-90}
\caption{Fits to the equatorial limb polarization spectrum compared to the measured spectrum (thin line). The thick line 
is a best fit for the spectral region 5300 - 7500\,\AA\ and the dotted line
for 7700 - 9300\,\AA.}
\label{6212fig5}
\end{figure}

The parameters $c_1$, $c_2$, and $c_3$ can then be determined with a least 
square procedure. Particularly good fits were obtained for the spectral
region from 5300\,\AA\ to 7500\,\AA\ 
(see Fig.~\ref{6212fig5}).

The best fit to the equatorial limb polarization for the wavelength region 
5300 - 7500\,\AA\ is:
\begin{equation}
p_Q(\lambda) [\%] =  8.4 -  9.1\cdot\lambda[\mu{\rm m}] -2.2\cdot A(\lambda).
\end{equation} 

The standard deviation of the fit is 
  less than 0.1\,\%. The largest deviations are
due to the telluric absorption bands at 6870\,\AA\ (B-band) and 7200-7300\,\AA\
  (water absorption) which were 
not corrected in the albedo spectrum. Due to the uncorrected A-band absorption 
at 7590\,\AA, the 
upper boundary of the least square fitting procedure was fixed to 7500\,\AA. 

The fit to the longer wavelength region 7700\,\AA\ to 9300\,\AA\ yields a
flatter wavelength dependence and a stronger albedo dependence 
($c_1 =6.0 $, $c_2 =-5.6 $ and $c_3=-4.5 $).

\subsubsection{Spatial sampling and full slit signal}\label{6212fullslit}

According to a simple analytic model for Rayleigh scattering atmospheres
(see e.g. paper {\sc i}; van de Hulst~\cite{vandehulst80}) the
limb polarization feature is expected to be a narrow, strongly peaked feature
along the limb. In our data this narrow peak is not resolved due to the
seeing limited ($\approx 1$~arcsec) resolution. Therefore, the
apparent strength of the limb polarization $p_Q(\lambda)$ depends strongly on
the not so well defined seeing conditions during the observations. In
addition we have to consider also the slit width which defines the
spatial region over which the intensity and Stokes fluxes 
are sampled. For Uranus we employed a narrow slit (0.5~arcsec) which
caused only a small degradation of the polarization signal. 

Choosing a radial bin for the quantification of the limb polarization
is an arbitrary procedure. For this reason it is
reasonable to employ the polarization signal integrated over
the entire slit to quantify the limb polarization. A fit 
according to Eq. (\ref{6212fitequation}) to the total slit 
spectropolarimetry of the equatorial slit as
given in Fig. \ref{6212fig3} yields the parameters 
$c_1=3.6$, $c_2=-3.8$, and $c_3=-1.3$ (again for the spectral 
region $5300 - 7500$\,\AA). 

In Sect. \ref{6212seeing} we derive a correction factor which accounts for the
seeing and the spatial sampling of our observations.

\section{Disk profiles for Uranus}

The spectropolarimetric data can be employed to investigate the
intensity and polarization profiles through the disk of Uranus for any wavelength interval covered by our observations, a sort of center to limb profiles.
Since the slit is
much longer than the planet's diameter our data provide profiles
roughly along the equator (equatorial slit) and profiles 
from the South pole to high northern latitudes (meridional slit).

Again, we focus our analysis on the
spectral regions C1 -- C5 (continuum / high albedo regions) and M1 -- M7
(strong methane bands) as defined in Table~\ref{6212regionsdef}. 

\begin{figure}[h!]
\centering
\epsfig{file=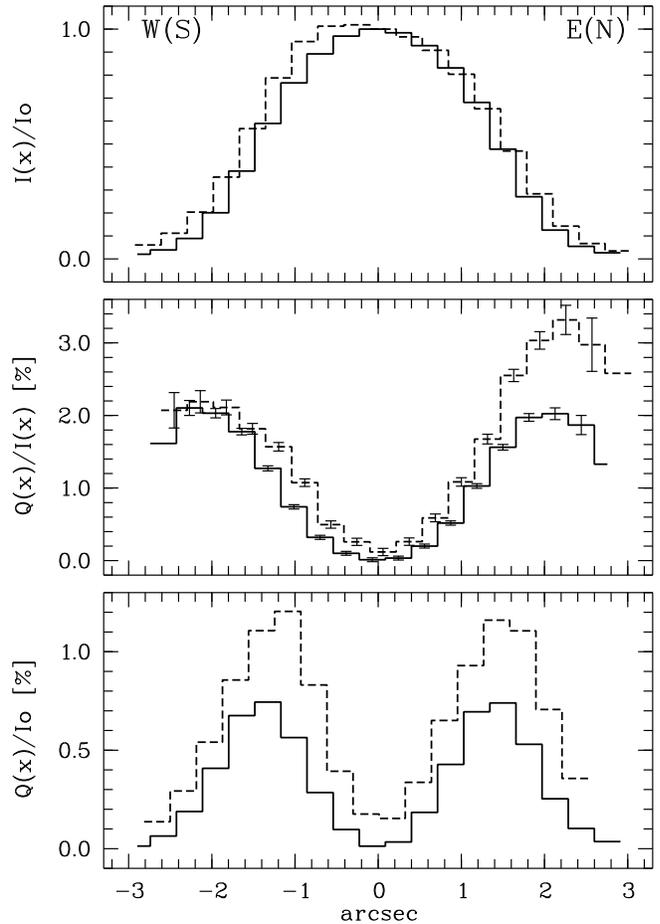,width=9cm}
\caption{Uranus equatorial (solid) and meridional (dashed) disk profiles for the 
continuum (high albedo)
  wavelength range 5827\,\AA\ to 5883\, \AA\ (C1). The panels give the flux
  $I(x)/I_0$ (top), the normalized Stokes parameter $Q(x)/I(x)$ (middle) and the Stokes
  flux $Q(x)/I_0$ (bottom).} 
\label{6212fig6}
\end{figure}

\begin{figure}[h!]
\centering
\epsfig{file=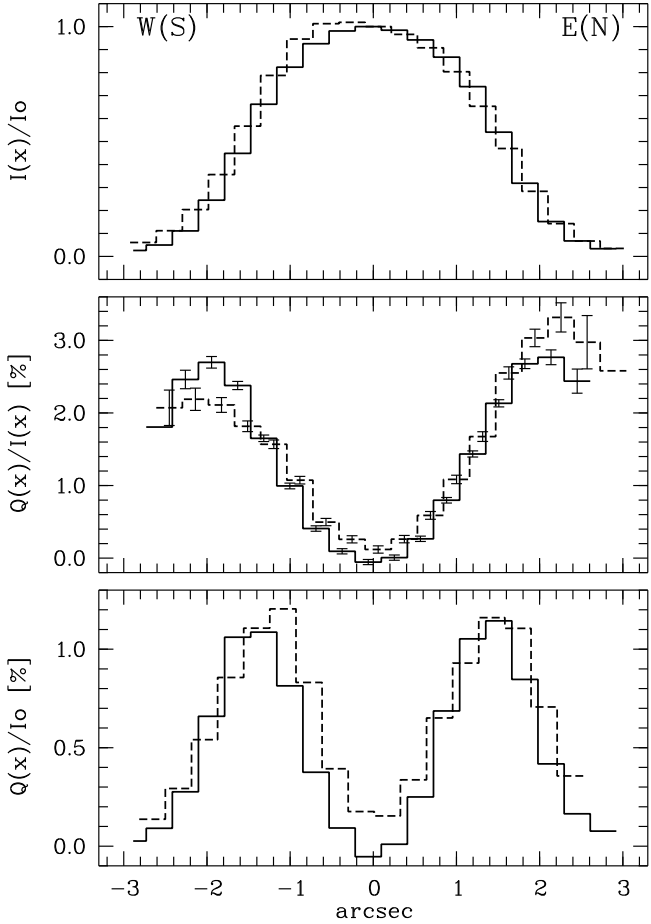,width=9cm}
\caption{Uranus equatorial (solid) and meridional (dashed) disk profiles for the 
methane band centered on 5760\,
  \AA\ (M2). The panels give the flux
  $I(x)/I_0$ (top), the normalized Stokes parameter $Q(x)/I(x)$ (middle) and the Stokes
  flux $Q(x)/I_0$ (bottom).}
\label{6212fig7}
\end{figure}

Figures \ref{6212fig6} and \ref{6212fig7} show examples 
for disk profiles for the C1-region around 5850\,\AA\ and the M2 methane band at
5760\,\AA. The flux of the disk profiles
was normalized at the disk center $I_0=I(x=0'')=1$ and also the
Stokes flux $Q$ is expressed relative to $I_0$.


\subsection{Disk profiles for the intensity}

The intensity profiles show some pronounced differences between the equatorial and the meridional directions as well as between different spectral regions. 
All equatorial profiles are essentially symmetric while
the meridional profiles are more or less asymmetric, showing a brighter
South pole when compared to the northern latitude (see
Fig.~\ref{6212fig7}). 
This asymmetry is particularly strong in the deep methane bands. In this section we further present a relationship between intensity profile structure and albedo for both slit orientations.

\begin{figure}[h]
\centering
\epsfig{file=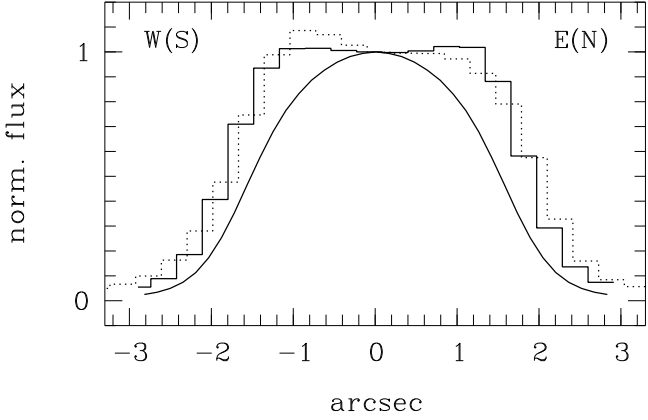,width=6.5cm,angle=-90}
\caption{Disk intensity profile for the methane absorption M5, showing the
strongest asymmetry between South and North (dotted line) and equatorial limb brightening (solid) effect.  
For comparison the symmetric reference profile $I_{\rm ref}$ is plotted in addition (Gaussian-like).}
\label{6212fig8}
\end{figure}

The equatorial cuts have a Gaussian-like profile for the
continuum / high albedo bands (Fig.~\ref{6212fig6}) and 
flat top profiles for the deep
absorptions (Figs.~\ref{6212fig7} and~\ref{6212fig8}). The flat top profiles 
can be explained as not well resolved limb brightening effect. 
  
These intensity features of Uranus are also clearly present in imaging studies
based on high resolution HST observations
(e.g. Karkoschka~\cite{karkoschka01}; Rages et al.~\cite{rages04}). Compared to the HST data,
we can only see the strongest spatial features due to the much
lower spatial resolution of our data. However, the 
spectropolarimetric data yield disk profiles for any spectral region
in the covered wavelength range and therefore our data are
particularly well suited to investigate the systematic behavior
of the asymmetry and the limb brightening for Uranus. 

\begin{table}[h!]
\center
\caption{Uranus meridional asymmetry parameter $\Delta_{mer}$ and limb
  brightening parameter $\Sigma_{eq}$ at the equatorial regions for spectral features 
  C1 -- C5 and M1 -- M7 
  as defined in Table~\ref{6212regionsdef}.}
\label{6212uranusquant} 
\begin{tabular}{ccc}
\noalign{\smallskip\hrule\smallskip}
feature & $\Delta_{mer}$ & $\Sigma_{eq}$ \cr

\noalign{\smallskip\hrule\smallskip}
C1  & $-$0.017  & $-$0.082 \cr
C2  & $-$0.030  & $-$0.019\cr
C3  & $-$0.008  & 0.019 \cr
C4  & 0.038 & 0.048\cr
C5  & 0.102 & 0.035\cr
\noalign{\smallskip\hrule\smallskip}
M1  & 0.031 & 0.070 \cr
M2  & 0.079 & 0.039 \cr
M3  & 0.095 & 0.375 \cr
M4  & 0.141 & 0.149 \cr
M5  & 0.160 & 0.529 \cr
M6  & 0.102 & 0.467 \cr 
M7  & 0.156 & 0.341 \cr 
\noalign{\smallskip\hrule\smallskip}
\end{tabular}
\end{table}

In order to quantify the asymmetry between South and North and the limb brightening
effect we construct first a symmetric disk profile as reference 
profile $I_{\rm ref}(x)$ for Uranus by averaging all 5 equatorial continuum 
profiles (C1 -- C5) and their mirrored profiles (mirrored at $x=0''$).
As next step, we determine the relative deviation $\delta(x)$ 
for the equatorial and meridional profiles from the reference profile by calculating

\begin{equation}
 \delta(x)={I(x)-I_{\rm ref}(x)\over I_{\rm ref}(x)}\,. 
\end{equation}

Due to our normalization, the deviation
in the middle of the profiles is for all our profiles 
$\delta(x=0'')=0$. Strong signals in $\delta(x)$ are obtained
for $x\approx \pm 1''$. We calculated the averages of the deviations 
$\langle \delta(x)\rangle$
for all four planetary orientations on the disk selecting profile 
regions around $x\approx 1''$. The selected regions correspond to the ``grey'' 
boxes in Fig.~\ref{6212fig1}, defined by:

\begin{eqnarray}
\delta_{W,S}= \langle \delta(x)\rangle & \quad{\rm for} & -1.41''<x<-0.47''
         \nonumber \\
\delta_{E,N} = \langle \delta(x)\rangle & \quad{\rm for} & +0.47''<x< +1.41''\,.
         \nonumber 
\end{eqnarray}

Based on this we calculated the meridional asymmetry parameter:
\begin{equation}
 \Delta_{mer} = \delta_{S}-\delta_{N}
\end{equation}
and the equatorial limb brightening parameter:
\begin{equation}
\Sigma_{eq} = \delta_{W}+\delta_{E}\,,
\end{equation}
which are given in Table~\ref{6212uranusquant}
for the wavelength intervals C1 -- C5 and M1 -- M7. Large values
indicate a relatively strong asymmetry between South and North or a relatively strong
limb brightening. Figure \ref{6212fig8} illustrates the asymmetry for the central meridian and the equatorial limb brightening for the most extreme case, the M5 absorption band 
at 7270\,\AA. 

It should be noted
that the absolute values of $\Delta_{mer}$ and $\Sigma_{eq}$ contain
little diagnostic information, because they depend on the seeing and
the size of the chosen spatial interval. 

\begin{figure}[h]
\centering
\epsfig{file=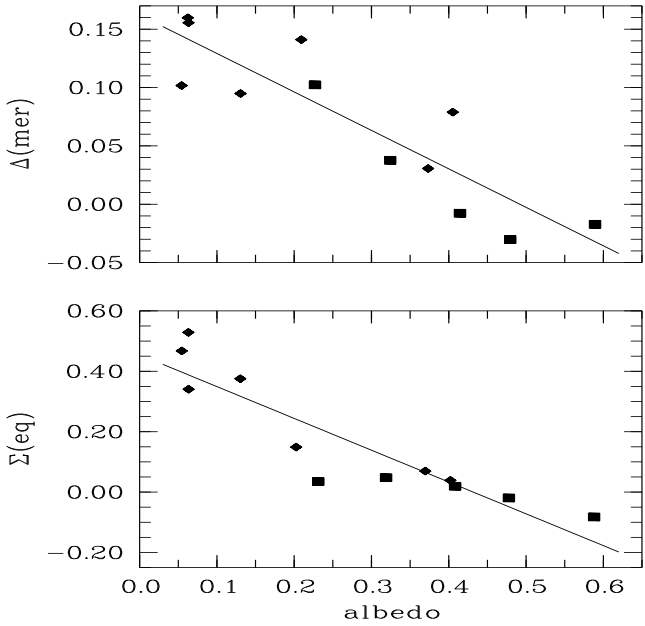,width=9cm, height=9cm}
\caption{Uranus meridional asymmetry parameter $\Delta_{mer}$ vs. albedo (top) and limb
  brightening $\Sigma_{eq}$ vs. albedo (bottom) for high albedo 
  spectral regions C1 -- C5 ($\square$) and methane absorption bands M1-M7 
  ($\diamond$).}
\label{6212fig9}
\end{figure}

More interesting information is obtained from the relative strength of the 
$\Delta_{mer}$ and $\Sigma_{eq}$ parameters for different wavelength bands. They are all affected in a similar way by the limited spatial resolution. 
Our data show a tight anti-correlation
between the meridional asymmetry parameter $\Delta_{mer}$ and albedo $A$ as well as between limb brightening 
$\Sigma_{eq}$ and $A$ which is illustrated in Fig.~\ref{6212fig9}.

The HST observations of Uranus described by Karkoschka~(\cite{karkoschka01})
also clearly show that the intensity asymmetry and limb brightening are
particularly strong in the methane bands. However, our long-slit spectroscopy
allows to select any particular wavelength band for a center-to-limb profile
analysis. To our knowledge, the dependence of the asymmetry and limb brightening
effect on the albedo has not been described previously. The asymmetry between South pole and northern latitudes is of course a seasonal effect of Uranus. 

To investigate the properties of the scattering layers in
Uranus and Neptune via the limb brightening effect, it would be desirable to have long-slit spectroscopy with a spatial resolution as offered by HST.

\subsection{Disk profiles for the polarization}

Disk profiles for the polarization are obtained for the
normalized Stokes parameter $Q(x)/I(x)$ (short $Q/I$) and the Stokes flux $Q(x)/I_{0}$ (short $Q$), where positive
values indicate a polarization parallel to the slit or perpendicular
to the limbs. A few examples are plotted in Figs. \ref{6212fig6} and
\ref{6212fig7}.

As already
described in the previous sections, the polarization is low
in the center of the disk and high at the limb at all wavelengths
of our spectrum. Typically $Q/I$ 
increases steadily with radius $r=|x|$, and reaches a constant
value around $r\approx 2''$, where the photon noise starts to
dominate the measurements. The Stokes flux $Q$ 
increases from $r=0''$ until about $r\approx \pm 1.5''$,
and decreases together with the intensity
profile further out to zero. This behavior is in agreement with the imaging  
observations described in paper {\sc i}.

For the equatorial slit observations all $Q/I$ and $Q$ disk profiles
are essentially symmetric, and they have roughly 
the same spatial structure (Figs.~\ref{6212fig6} and
\ref{6212fig7}). Only the strength of the polarization signal shows a clear wavelength dependence. 

For the meridional slit there is an obvious asymmetry in the $Q/I$ polarization
signal. For most spectral regions the $Q/I$ polarization is typically 
30\,\% higher at the northern limb than at the South limb (Tab.~\ref{6212regionsdef}, 
Figs.~\ref{6212fig6} 
and \ref{6212fig7}). The asymmetry is particularly strong
for the inter band regions $C4$ and $C5$. This effect is also described
in Sect.~\ref{6212sectEW}, where the spectropolarimetric signal for the southern and northern limb is compared.

Surprisingly, the $Q$-flux profiles in meridional direction are rather symmetric 
(Figs.~\ref{6212fig6} and \ref{6212fig7}). Thus, 
the higher intensity $I$ at the South pole, combined with the lower 
normalized polarization $Q/I$, yield together a Stokes flux $Q$ at a similar 
level as for the darker, but higher polarized northern latitudes. 
Whether this is a fortuitous coincidence or an indication of a planet-wide homogeneous haze layer producing everywhere more or less the same limb polarization flux remains to be
investigated.

\subsubsection{Seeing corrected limb polarization parameters for Uranus} \label{6212seeing}

For comparison with models and future observations of Uranus, it appears
useful to derive well defined parameters for the strength of the 
limb polarization. In particular, such values should be corrected for
the polarization cancellation introduced by the seeing, and the
averaging effects due to the finite slit width.

For this reason we first determine the flux weighted
mean polarization for a line (slit) through the center of the planet
$\langle Q/I \rangle_{\rm line}$. This is calculated
according to $\langle Q/I \rangle_{\rm line}=\Sigma Q/\Sigma I$, where
$\Sigma Q$ and $\Sigma I$ are the polarization flux $Q(x)$ and flux $I(x)$, summed
up along the entire slit (identical to the procedure in paper {\sc i}).
The slit integrated polarization $\langle Q/I \rangle_{\rm line}$ is
of course significantly lower than the values derived for a small limb region
given in Table \ref{6212regionsdef} or shown in Fig.~\ref{6212fig3} for the equatorial slit. 

The seeing introduced by the Earth's atmosphere causes a polarization cancellation
effect. As explained in detail in paper {\sc i}, the opposite sky polarization
components, e.g. the $+Q_{\rm sky}$ components at the equatorial limbs, and 
the $-Q_{\rm sky}$ components at the northern and southern limbs, overlap and cancel each other
if the seeing is substantial, say seeing $\gtrsim 0.5~{\rm R}_{\rm planet}$. 
A second effect of the seeing is that the signal from regions adjacent to
the slit spill into the slit. This averages down the measurable
limb polarization. In addition, there is also a reduction of the
integrated limb polarization due to the finite slit width. For Uranus this is
a small effect, because the data were obtained with a very narrow slit of 
0.5$''$. 

For our observations the
seeing-limited resolution was determined to be 
1.09$''$ at 6000\,\AA\ and 1.00$''$ at 8000\,\AA\
taken from observations of standard stars observed with the same setup. 
However, the seeing conditions can change on short time scales, 
introducing uncertainties on the order of $\pm 0.1''$ to $\pm 0.2''$.

This effect of polarization cancellation due to the limited resolution 
was modeled in paper {\sc i}, and seeing correction factors for 
the imaging polarimetry were determined. We have repeated these
calculations for observations through
a 0.5$''$ wide slit and obtained essentially the same 
seeing dependence as for the imaging polarimetry (see Fig.~11 in
paper {\sc i}). From these simulations we found that the measurable polarization
is reduced by a factor 0.96 due to the width (0.5$''$) of the slit
and a factor of 0.76 due to the seeing of $1''$ when compared to
the intrinsic limb polarization integrated along a slit through the
disk center. 

Thus, in Table~\ref{6212regionsdef2} we give corrected values for the intrinsic limb polarization $\langle Q/I \rangle _{\rm line} ^{\rm corr}$ for Uranus by multiplying the measured values $\langle Q/I \rangle _{\rm line}$
with a factor of 1.37. The uncertainty in this correction factor is
about $\pm 0.15$ mainly due to the not precisely defined seeing. 
For the intrinsic limb polarization other error sources like photon noise
or the calibration of the polarization data are less important. 

\begin{table}[h!]
\center
\caption{Line (slit) integrated, seeing corrected polarization $\langle Q/I \rangle_{\rm line} ^{\rm corr}$
  for the equatorial and meridional slits through the center of Uranus for the selected
  spectral features (Table~\ref{6212regionsdef}). The listed values are corrected (measured
  values $\times 1.37$; see text) for the seeing and the employed slit width
  of our observations.}
\label{6212regionsdef2}
\begin{tabular}{cccc}
\noalign{\smallskip\hrule\smallskip}
feature & \multicolumn{3}{c}{$\langle Q/I \rangle_{\rm line} ^{\rm corr}$ [\%]} \cr
        &equatorial &  South &  North \cr
\noalign{\smallskip\hrule\smallskip}
C1 & 0.90 & 1.02 & 1.19\cr
C2 & 0.80 & 0.99 & 1.13\cr
C3 & 0.66 & 0.90 & 1.05\cr
C4 & 0.41 & 0.57 & 0.89\cr
C5 & 0.18 & 0.24 & 0.67\cr
\noalign{\smallskip\hrule\smallskip}
M1 & 1.47 & 1.46 & 1.85\cr
M2 & 1.26 & 1.40 & 1.55\cr
M3 & 1.61 & 1.46 & 1.57\cr
M4 & 1.01 & 1.13 & 1.28\cr
M5 & 1.15 & 0.95 & 1.04\cr
M6 & 0.88 & 0.75 & 1.03\cr
M7 & 0.71 & 0.46 & 1.08\cr
\noalign{\smallskip\hrule\smallskip}
\end{tabular}
\noindent
\end{table}

The same correction factor can also be applied to the fit to the equatorial slit-integrated spectropolarimetric signal determined in Sect.~\ref{6212fullslit}. 
Thus, the corrected fit for the intrinsic equatorial full slit polarization for
the spectral region 5300 to 7500\,\AA\ can be given by:
\begin{equation}
\label{6212equationfitpq}
p_Q(\lambda) = 4.9 - 5.2 \lambda\,[\mu{\rm m}] - 1.8 A(\lambda)\,.
\end{equation}

\section{Spectropolarimetric structure of Neptune}\label{6212specpolneptune}
\begin{figure}[h]
\centering
\epsfig{file=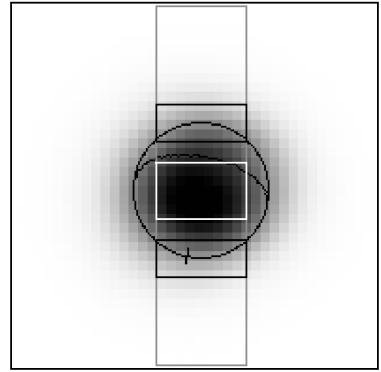,width=5cm}
\caption{Intensity image of Neptune in the i-band with North-South slit position. North is up and East is left. The South pole is marked with a dash. The grey stripe indicates the position of the
slit. The black boxes show the ``limb'' area and the white box 
the ``central part''. At the time of our observations the diameter was 2.24$''$, the phase angle 1.7$^{\circ}$, the position angle of the South pole 169$^{\circ}$ and the distance of the South pole from the center of the disc was 0.97$''$.}
\label{6212fig10}
\end{figure}

Spectropolarimetry of Neptune was obtained with a North-South oriented slit and almost
the same instrument set-up as for Uranus. The only difference
was the slit width of 1.5$''$, which was  much broader than for Uranus. 
We also define three slit regions for Neptune, a North limb and South limb section
of 0.625$''$ each, a central region of 0.942$''$ and an integrated ``total slit'' of 4.71$''$ length (Fig.~\ref{6212fig10}).

At the time of our observations Neptune had a diameter of 2.24$''$. Compared
to this size the width of the slit of 1.5$''$ and the seeing conditions 
of about $1''$ are large and not well suited to resolve spatial features
on the planetary disk. Almost the entire planet was contained in the slit
except for the eastern and western limbs. With higher spatial resolution observations it was shown that Neptune displays also a limb brightening effect similar to Uranus (e.g. Hammel et al. \cite{hammel89}; Baines \& Hammel \cite{baines94}).

\begin{figure}[h]
\centering
\epsfig{file=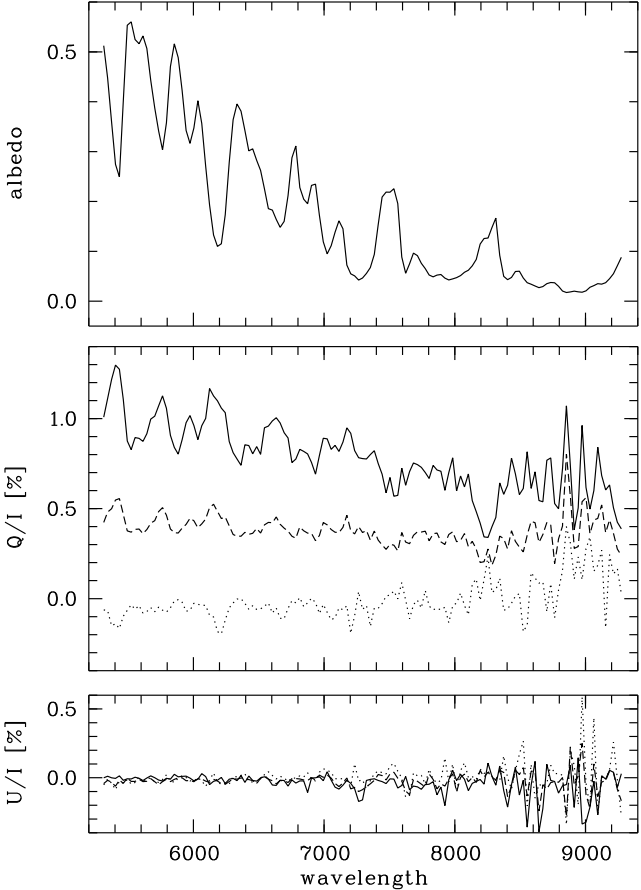,width=9cm}
\caption{Spectropolarimetry of Neptune for the slit in North-South direction.
Top panel: geometric albedo spectrum $A(\lambda)$.
Middle panel: normalized Stokes $Q/I$ polarization spectra for the average of
the North and South limb regions (solid line), the central part (dotted line) and the
entire slit (dashed line). Bottom panel: Same as the middle panel but for Stokes $U/I$.}
\label{6212fig11}
\end{figure}

Despite the bad resolution of our Neptune observations a clear 
polarization signal is detected. This is shown in 
Fig.~\ref{6212fig11}, which presents the spectropolarimetry of Neptune in a
similar form as for Uranus (Fig.~\ref{6212fig3}). The albedo spectrum 
$A(\lambda)$ was constructed with a normalization to the Neptune albedo 
spectrum of
Karkoschka (\cite{karkoschka94}). Spectroscopically, Neptune is very similar to Uranus (see Karkoschka~\cite{karkoschka98}).

The spectropolarimetric signal of Neptune for the average of the northern and southern limbs 
is qualitatively also very similar to Uranus (Fig~\ref{6212fig11}). The 
$Q/I$ polarization spectrum is high and positive at the limbs, or parallel to the slit, but is low in the disk center, and essentially zero for the $U$-direction.   
$Q/I$ shows an overall decrease in polarization towards longer
wavelength and an enhanced polarization in the strong methane bands, when
compared to the adjacent continuum or higher albedo regions.   
The level of the measured limb polarization in Neptune is significantly 
lower than for Uranus because the planetary disk is less well spatially resolved. 

The spectropolarimetry for the central region shows a weak and negative Q/I polarization
spectrum. This can be easily understood as polarization signal from the eastern and western limb which spills into the central slit region. As the eastern and western limb polarizations are oriented perpendicular to the limb, they produce a
perpendicular (negative) signal for the North-South slit. The slit integrated
polarization is small because of the small spatial resolution. For spatially
unresolved observations with a slit which is 
wider than Neptune, a total net polarization close to zero is expected, because the
centro-symmetric polarization structure of Neptune (see paper {\sc i}) would
produce an almost perfect cancellation of positive and negative polarization features.   

Parameters for the limb polarization of Neptune are given in Table~\ref{6212regionsdefnept}.
Because Neptune is spectroscopically very similar to Uranus, the same spectral
features were selected. The mean albedo for the spectral intervals was
determined from the albedo spectrum $A(\lambda)$. The limb polarization is
only given for the average of the northern and southern limb regions. The two measurements
are averaged because no significant difference is detectable between the North and the South limb. 

Because the spatial resolution of our observation is poor, it is not
possible to deduce reliable parameters for the intrinsic
limb polarization of Neptune. However, the imaging polarimetry presented
in paper {\sc i} indicates that the intrinsic limb polarization of Neptune
is about 1.5 times higher than that of Uranus.  
  
\begin{table}[h!]
\caption{Limb polarization $\langle Q/I \rangle$ (in [\%]) for Neptune for the average
  of the North and South limbs, $\langle Q/I \rangle _{NS}$ and for the total meridional slit $\langle Q/I \rangle _{t}$, for selected wavelength intervals. Column 2 gives the averaged
albedo $A(\lambda)$ for the corresponding interval.}
\label{6212regionsdefnept}
\centering
\begin{tabular}{cccc}
\noalign{\smallskip\hrule\smallskip}
feature  & albedo & $\langle Q/I \rangle _{NS}$  & $\langle Q/I \rangle _{t}$\cr
\noalign{\smallskip\hrule\smallskip}
C1  & 0.49 & 0.85 & 0.37\cr
C2  & 0.37 & 0.80 & 0.36\cr
C3  & 0.30 & 0.77 & 0.34\cr
C4  & 0.22 & 0.62 & 0.29\cr
C5  & 0.14 & 0.38 & 0.23\cr
\noalign{\smallskip\hrule\smallskip}
M1  & 0.32 & 1.22 & 0.52\cr
M2  & 0.33 & 1.09 & 0.48\cr
M3  & 0.11 & 1.08 & 0.47\cr
M4  & 0.16 & 0.97 & 0.42\cr
M5  & 0.06 & 0.83 & 0.38\cr
M6  & 0.05 & 0.67 & 0.33\cr
M7  & 0.05 & 0.67 & 0.33\cr
\noalign{\smallskip\hrule\smallskip}
\end{tabular}
\end{table}

\subsection{Fitting the polarization spectrum of Neptune}

We present also a fit to the limb polarization spectrum of Neptune for a
quantitative description. The same procedure is applied as described in
Sect.~\ref{6212sectfit}. This yields for Neptune for the average of the North and South limb polarization spectrum and the
wavelength range of 5300 - 7500\,\AA\ the fit:
\begin{equation}
\label{fitequationneptune}
p_{Q}(\lambda)[\%] = 3.35 - 3.35\cdot\lambda[\mu m] - 1.10\cdot A(\lambda).
\end{equation}
It is interesting to note that the fit parameters for Neptune behave like 
$c_1\approx -c_2$ and $c_1\approx -3 c_3$. These are almost identical 
proportionalities as derived for the limb polarization in Uranus. This
indicates that the wavelength and albedo dependence of the limb polarization
is very similar for Neptune and Uranus. 
 
For the red end of the North-South spectropolarimetry (7700 - 9300\,\AA) the derived
parameters are $c_{1} = 1.75$, $c_{2} = -1.15$ and $c_{3} =-2.75$. For the
longer wavelength region the wavelength dependence is flatter and the albedo
dependence stronger. 

\begin{figure}[h]
\centering
\epsfig{file=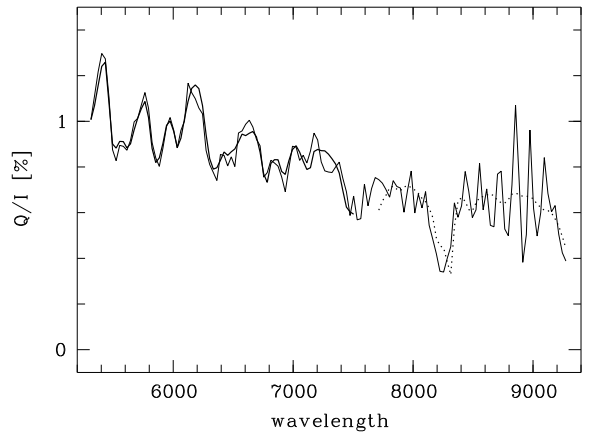,angle=270,width=9cm} 
\caption{Fits to the North-South limb polarization spectrum of Neptune compared to the measured spectrum (thin
  line). The thick line is a fit for the spectral region 5300 - 7500\,\AA, 
and the dotted line for 7700 - 9300\,\AA.}
\label{6212fig12}
\end{figure}

\section{Discussion and Conclusions}\label{6212discussion}
\subsection{Comparison with the imaging polarimetry of paper\,\sc{i}}

The R, i and z-band imaging polarimetry of Uranus and Neptune reported
in paper {\sc i} reveals the same overall polarization behavior as
the spectropolarimetric data presented in this work. Both, imaging polarimetry
and long-slit spectropolarimetry, show for Uranus and Neptune essentially 
no polarization in the disk center and a strong limb polarization with a position angle 
perpendicular to the limb. 

Already the polarimetric images taken in a few, mainly broad-band, filters 
define a general trend, that the limb polarization is lower for longer wavelengths. 
However, the wavelength dependence of the limb polarization is described
in much more detail with the long-slit spectropolarimetry presented in this paper. From spectropolarimetry
it is obvious that there is not only a decrease in the limb polarization 
with wavelength but also a very tight anti-correlation with the albedo. 

Our spectropolarimetry and the imaging polarimetry of Uranus and Neptune yield
further results which are complementary. 

Very interesting is the spectropolarimetric
result that the relatively dark northern latitudes have
a higher level of polarization $Q/I$ when compared to the other limbs. With
the imaging polarimetry it was not possible to find this result due to the
delicate alignment requirements for the images with opposite polarization in
the data reduction process. 
 
The imaging polarimetry was taken under better seeing
conditions (0.8$''$) and the spatial resolution was not further degraded
by a wide slit, as it was for Neptune in our spectropolarimetry. Therefore, the
imaging polarimetry allowed to derive a seeing corrected  
limb polarization average for both planets, indicating that the intrinsic
limb polarization for Neptune is about a factor of $\approx$ 1.5 higher than
for Uranus.     
 
A good quality test of the data analysis provides the comparison of
the derived intrinsic limb polarization of Uranus from imaging polarimetry
and spectropolarimetry. For this we have to convert the derived values
for the disk averaged limb polarization $\langle Q_r/I \rangle$ from imaging
polarimetry into the slit (line) averaged limb polarization $\langle Q/I
\rangle_{\rm line}$ as obtained from the spectropolarimetric data. 
This conversion depends somewhat on the radial dependence of the intensity and the
limb polarization. However, the simple model results presented in paper {\sc
  i} indicate that a conversion relation $\langle Q_r/I \rangle \approx 1.6\, \langle Q/I
\rangle_{\rm line}$ is adequate for the expected center to limb profiles.  

Thus, the converted parameters for the intrinsic limb polarization of 
Uranus from the filter polarization is $\langle Q/I \rangle_{\rm line} ^{\rm corr}=
0.53\,\%$ for the $\lambda_{673}$-filter (albedo=0.32), 0.59\,\% for the
$\lambda_{729}$-filter, corresponding to M5, (albedo=0.05), 0.46\,\% for the i-filter (albedo=0.19) and 0.27\,\% for the
z-filter (albedo=0.12).

\begin{figure}[h]
\centering
\epsfig{file=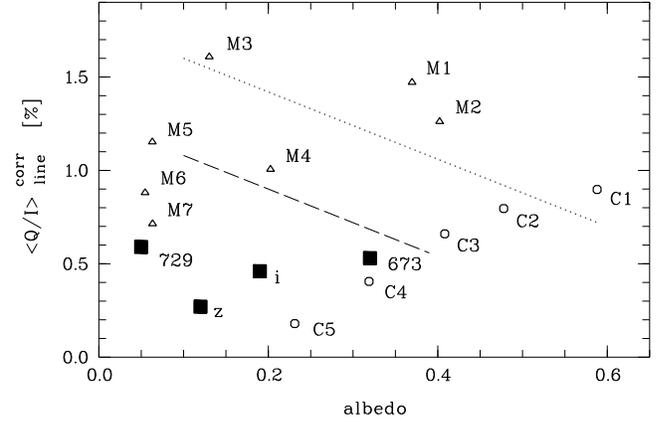,angle=270,width=9cm} 
\caption{Seeing corrected, full slit polarization for the equatorial slit of Uranus vs. albedo from Table \ref{6212regionsdef2}. The circles are the high albedo regions (C1 -- C5) and the triangles are the methane absorption bands (M1 -- M7). The dotted line gives the full slit polarization at 6000\,\AA, and the dashed line, at 7000\,\AA, according to Eq. (\ref{6212equationfitpq}). The filled squares denote the measured polarization with the four filters from paper {\sc i} (broad band filters i and z and the narrow band filters centered at 673 and at 729~nm).}
\label{6212fig13}
\end{figure}

In Fig. \ref{6212fig13} the limb polarization $\langle Q/I \rangle _{\rm line} ^{\rm corr}$ and the albedo from the filter polarimetry are compared with  the spectropolarimetry. The broad band filters i and z lie between the values for the reddest wavelength intervals C5 and M6 and M7 as expected for a broad band average. Also for the narrow band filter $\lambda _{673}$ located in wavelength between M4 and C3  the agreement is quite good. Only for the $\lambda _{729}$ filter the measured polarization from the filter polarimetry is significantly lower than the spectropolarimetric results (M5). As described in paper {\sc i}, the accuracy of the measurement in this filter is particularly low due to the low signal-to-noise. Thus, it may be concluded that the agreement between imaging polarimetry and spectropolarimetry is good, except for the low quality  $\lambda _{729}$-filter observations.

This gives confidence that the analysis and the applied seeing corrections are consistent for the two data sets. 

\subsection{Limb polarization and atmosphere models}
 
The properties of the limb polarization for Uranus and Neptune can be 
summarized as follows: higher polarization for shorter wavelength, enhanced
polarization in the methane bands when compared to the continuum or 
inter band regions, and in Uranus a higher continuum/inter band polarization 
for the darker northern latitudes when compared to the bright South Pole. 

An enhanced polarization in the strong methane bands is also observed 
for Jupiter (Wolstencroft \& Smith \cite{wolstencroft78}; Joos et al. \cite{joos05}). Thus, it seems 
that the tight anti-correlation between limb polarization and geometric
albedo (within a small wavelength domain) is a generic 
feature of gas planets with a substantial amount of reflected light due to
Rayleigh type scattering. 

According to simple Rayleigh scattering models a limb polarization is 
expected for scattering atmospheres with some amount of Rayleigh-type 
scattering. 

However, for homogeneous (semi-infinite) Rayleigh scattering
atmospheres one does not expect an enhanced limb polarization
in deep absorption bands (see Table~3 in paper {\sc i}) except for 
a small range of single scattering albedos around 0.8 and 1.0 corresponding to geometric 
albedos between 0.3 to 0.79. A tight anti-correlation between 
polarization and albedo is obtained for models with 
finite Rayleigh scattering layers above diffusely reflecting 
ground layers or deeper layers with different albedos 
(see Table~4 in paper {\sc i}). This can 
easily be understood as a constant amount of Rayleigh scattered radiation 
producing the limb polarization, combined with an albedo dependent 
contribution of unpolarized light from diffusely reflecting lower layers. 

The limb brightening effect points to a similar stratification in the atmosphere of 
Uranus and Neptune. As predicted by Belton \& Price (\cite{belton73}) and 
illustrated by e.g. Hammel et al.~(\cite{hammel89}) or Sromovsky 
(\cite{sromovsky05a}), only inhomogeneous scattering atmospheres with 
a single scattering opacity increasing with depth will show a limb 
brightening. 

It is well known that Uranus and Neptune have optically thin scattering 
haze layers high in the atmosphere. This finding is based on careful 
analyses and modeling of the albedo spectrum, center to limb profiles, 
and other multi-wavelength studies of these planets
(e.g. Baines \& Bergstralh \cite{baines86};  Hammel et al.~\cite{hammel89}; 
Baines \& Smith \cite{baines90}; Baines \& Hammel \cite{baines94}; 
Sromovsky \cite{sromovsky05a}). 

With the newly detected limb polarization of Uranus and Neptune it 
would now be possible to test the existing models with radiative transfer
calculations including polarization. It should not be expected that
the atmospheric models have to be radically changed for Uranus and Neptune
due to the polarization signal. However, one may expect that an analysis
including the scattering polarization would provide important new insights on 
the scattering properties for the aerosol particles in the haze layers of
these planets.

\clearpage

\begin{acknowledgements} 
We are indebted to the ESO La Silla support team at the 3.6m telescope
who were most helpful with our very special EFOSC2 instrument setup. We are
particularly grateful to Oliver Hainaut. 
This work was supported by a grant from the Swiss National Science
Foundation.
\end{acknowledgements}

\end{document}